\documentclass[aps,floats,epsf,twocolumn]{revtex4-1}
\usepackage{amsmath}
\usepackage{amsfonts}
\usepackage{amssymb}
\usepackage{color}
\usepackage{MnSymbol,wasysym}
\usepackage{ dsfont }
\newcommand{\beq}{\begin{equation}}
\newcommand{\eeq}{\end{equation}}
\newcommand{\beqa}{\begin{eqnarray}}
\newcommand{\eeqa}{\end{eqnarray}}

\usepackage{graphicx,amssymb,amsmath,subfigure}
\usepackage{bm}

\newcommand{\braket}[1]{\left<#1\right>}

\newcommand{\dd}{\mathrm{d}}
\newcommand{\ii}{\mathrm{i}}

\newcommand{\el}{\text{el}}

\begin{document}
\title{Theory of the Strain-Induced Magnetoelectric Effect in Planar Dirac Systems}

\author{Pablo Rodriguez-Lopez}
\email{pablo.rodriguez@csic.es}
\affiliation{Materials Science Factory, Instituto de Ciencia de Materiales de Madrid, CSIC, Cantoblanco; 28049 Madrid, Spain.}
\affiliation{GISC-Grupo Interdisciplinar de Sistemas Complejos, 28040 Madrid, Spain}
\author{Alberto Cortijo}
\email{alberto.cortijo@csic.es}
\affiliation{Instituto de Ciencia de Materiales de Madrid,
CSIC, Cantoblanco; 28049 Madrid, Spain.}

\begin{abstract}
The magnetoelectric response in inversion-breaking two dimensional Dirac systems induced by strain is analyzed. It is shown that, in the same way that the piezoelectric response in these materials is related to the valley Chern number, the strain-induced magnetoelectric effect is related both to the non trivial Berry curvature and the derivative of the orbital magnetic moment per valley. This phenomenon allows to locally induce and control charge densities by an external magnetic field in strained zones of the sample.
\end{abstract}

\maketitle
\section{Introduction}
\label{sec:intro}
The magnetoelectric effect, that is, the generation of an electric polarization $\bm{P}$ by an external magnetic field $\bm{B}$, or conversely, the generation of a magnetization $\bm{M}$ as a response to an external electric field $\bm{E}$. This phenomenon has received renewed interest in the Condensed Matter Physics community in the recent years due to the discovery of the Topological magnetoelectric effect in three dimensional topological insulators\cite{QTZ08,EMV09}, and, more importantly, in multiferroics\cite{SF05}, where a ferroelectric phase transition takes place when applying an external magnetic field\cite{KGS03} (or the ferromagnetic phase is controlled by an electric field\cite{LLA04}).
A very appealing property of multiferroic materials is that there is not only a coupling between the electric and magnetic degrees of freedom leading to this magnetoelectric effect, but also both degrees of freedom can be entangled with the elastic properties of the material, opening new venues for technological applications\cite{F05}.

Quite interestingly, a new type of magnetoelectricity induced by strain has been experimentally reported in transition metal dichalcogenides (TMDCs)\cite{LWX17}. TMDCs are layered compounds that can be synthesized or exfoliated down to the single layer. These compounds posses a hexagonal structure lacking any inversion center,  and the metallic atom typically possesses a large intra-atomic spin-orbit (SO) coupling. These two facts are responsible for the spin-valley coupling in these systems\cite{XLF12,MHS12,YWL14} and the presence of an orbital magnetic moment per valley\cite{YXN08}. Time reversal symmetry enforces the orbital magnetic moment per valley to be exactly opposite, cancelling each other in equilibrium.

On the other hand, in these materials, like in graphene, the symmetries of the lattice allow for a coupling between elasticity and electrons in the form of an effective vector field that enters as a Peierls substitution with opposite sign for each valley\cite{RRC15,ACJ16}. 

From this point of view, TMDCs (and in general, other two dimensional systems like graphene and boron nitride) display piezoelectricity, which is the generation of an electric polarization in systems due to the effect of an applied stress, or strain. If the strain depends on time, this polarization gives rise to net electric currents, while if the strain is not homogeneous, a charge density redistribution takes place. The inverse effect also occurs: under the effect of an external electric current, the system deforms to accommodate the induced stress. Due to the resemblance with the coupling with real electromagnetic fields, many topological phenomena related to electromagnetism can be associated to the elasticity as well, with the appropriate translation\cite{CFL15,CFL16}. This is the case of piezoelectricity, that, using the language of effective field theories, can be understood from the presence of a mixed Chern-Simons term \cite{VAA13} coupling the electromagnetic field $e\bm{A}$ and the elastic vector field $\bm{\mathcal{A}}^{\el}[u_{ij}]$, where $u_{ij}$ is the strain tensor. The opposite sign of the Berry curvature at each valley cancels out with the opposite sign of the coupling to the elastic vector field $\bm{\mathcal{A}}^{\el}$, giving a non vanishing response in time reversal invariant systems. Piezoelectricity has been experimentally observed in MoS$_2$\cite{WWY14,ZWX15}.

Returning to the magnetoelectric effect induced by strain in TMDCs, the physical explanation of the experimental observations given in \cite{LWX17} is related to the piezoelectric effect: The simultaneous presence of an external electric field $\bm{E}$ (or a current $\bm{J}$ related to $\bm{E}$ by the resistivity tensor), and a pseudo electric field, $\bm{\mathcal{E}}^{\el}\sim\dot{\bm{\mathcal{A}}}^{\el}$ (in virtue of the piezoelectric effect described for TMDCs), gives rise to a pseudo magnetic field $\bm{J}\times\bm{\mathcal{E}}^{\el}$ that generates a nonzero average magnetization $\bm{M}$. In order to trigger this $\bm{M}$, it is indispensable that $\bm{\mathcal{E}}^{\el}$ couples with opposite sign to both valleys.

However, it is important to notice that static and homogeneous strain configurations do not produce any current or polarization through the piezoelectric effect\cite{VKG10,JMV13}. This immediately means that, in order to explain the observations done in \cite{LWX17}, any possible mechanism cannot resort on the presence of piezoelectric currents or the presence of a strain induced pseudoelectric field. Using the electromagnetic analogy, a time independent and homogeneous strain produces a vector that acts as a pure gauge, that has no effect on the transport properties of the system. 

Motivated by this observation, in the present paper we put the strain-induced magnetoelectric effect in Dirac systems on solid grounds by analyzing the dual to this magnetoelectric effect: we describe how the presence of an external magnetic field $\bm{B}$ induces an electric polarization $\bm{P}$ in presence of elastic deformations described by $u_{ij}$. By means of the standard constitutive relations, the obtained components of the magnetoelectric tensor $\alpha_{ij}$ also serve to describe the generation of a magnetization when an external electric field (or electrical current) is applied together with the presence of an external strain.

\section{Generalities and Models}
\label{sec:models}
In general, the free energy density $F$ of a material under the effect of external electromagnetic fields can be written as
\beqa
F[\bm{E},\bm{B}]&=&-\epsilon_{ij}E_i E_j-\mu^{-1}_{ij}B_i B_j-\alpha_{ij}E_{i}B_{j}-...\label{freeenergy}
\eeqa
The macroscopic polarization $\bm{P}$ is obtained by varying $F$ with respect to the electric field
\beq
P_i=-\frac{\delta F}{\delta E_i}=\epsilon_{ij}E_j+\alpha_{ij}B_{j}.
\eeq

In the same way, we can define the magnetization $\bm{M}$ as
\beq
M_i=-\frac{\delta F}{\delta B_i}=\mu^{-1}_{ij}B_j+\alpha^{T}_{ij}E_{j},
\eeq
where $T$ means here the matrix transpose. Both quantities, $\bm{P}$ and $\bm{M}$, can be obtained from the same tensor $\alpha_{ij}$. So, our plan is to compute the dependence of $\bm{P}$ with $\bm{B}$, and automatically obtain the dependence of $\bm{M}$ with $\bm{E}$, which is the magnitude measured in the experiment. The dimensionality of the system and the absence of an explicit Zeeman term considerably reduce the number of non-zero components of $\alpha_{ij}$. We expect couplings only between $B_3$ and $E_1$ and $E_2$.

Concerning the microscopic details, we will consider as test bench  a system of two dimensional massive Dirac fermions related by time reversal symmetry, but not preserving inversion symmetry (non-centrosymmetric systems). For systems with hexagonal lattices, the low-energy, long wavelength momentum effective Hamiltonian around the so called $\bm{K}$ points can be written as

\begin{align}
H^{\eta} = H(\eta\bm{K}+\bm{k}) & =\eta v\sigma_{1} k_{1} + v\sigma_{2} k_{2} + \sigma_{3} m(\bm{k}).\label{Dirac1}
\end{align}

The parameter $\eta=\pm1$ stands for the valley degree of freedom. In the present work we will restrict outselves to a constant mass term $m(\bm{k})=m$, but the calculations can be done with $m(\bm{k})=m-\beta\bm{k}^2$. 

This simple situation describes Boron Nitride (BN)\cite{DBP16}, graphene under inversion breaking effects or the Kane-Mele model by considering the spin-orbit effect\cite{KM05}, bilayer graphene under the effect of an external perpendicular electric field\cite{CNM07}, and 2D materials of the graphene family: Silicene, Germanene and Stanene \cite{MO15}. In the case of TMDCs, we need to slightly modify the Hamiltonian (\ref{Dirac1}). In this case the spin plays a role, and the Hamiltonian that is compatible with time reversal symmetry reads

\begin{align}
H^{\eta}_{s}(\bm{k}) & = s\eta\lambda_{so}\sigma_{0} + \eta v\sigma_{1}k_{1} + \sigma_{2}k_{2} + (m - s\eta\lambda_{so})\sigma_{3},\label{DiracTMDCs}
\end{align}
where $\lambda_{so}$ is the effective spin-orbit coupling, and $s=\pm1$ is the spin projection\cite{XLF12}. For simplicity, we will neglect quadratic terms in momentum, and that the conduction bands are degenerate in energy.

While the coupling to a standard electromagnetic vector field $\bm{A}$ is dictated by the Peierls substitution, $\bm{k}\to\bm{k}-e\bm{A}$, for these elastic vector fields we have the effective substitution $\bm{k}\to \bm{k} - \eta\bm{\mathcal{A}}^{\el}$ (remember that $\eta$ stands for the valley degree of freedom), where $\bm{\mathcal{A}}^{\el}[u_{ij}]$ takes the form, for systems having $\mathcal{C}_3$ symmetry:
\begin{subequations}\label{Definition_Gauge_Ael}
\beq
    \mathcal{A}^{\el}_1=\frac{\beta}{a_0}(u_{11}-u_{22}),
\eeq
\beq
    \mathcal{A}^{\el}_2=-2\frac{\beta}{a_0}(u_{12}),
\eeq
\label{Afield}
\end{subequations}
where $u_{ij}$ is the strain tensor, $\beta$ is a dimensionless parameter that encodes the information of how the electronic structure changes under elastic deformations. The parameter $a_0$ is the lattice spacing.

%
\section{Kinetic approach}
\label{sec:kinetic}
To give a qualitative understanding of the role of the orbital magnetic moment and the Berry curvature in the previous results we will compute the induced current (or polarization) within a kinetic approach.

The kinetic approach resorts on the idea that at high enough Fermi level (or very short wavenunbers) the dynamics of the system is described by a distribution function in the phase space and a classical dynamics in terms of well defined trajectories and momenta\cite{XCN10}. In two dimensions, the equations of motion (EoM) read, for a given band: 
\begin{subequations}
\beq
D^{\eta}\dot{\bm{k}} = \bm{E}^{\eta} + e\bm{v}^{\eta}\times\bm{B},\label{eom1}
\eeq
\beq
D^{\eta}\dot{\bm{x}} = \bm{v}^{\eta} + \bm{E}^{\eta}\times\bm{\Omega}^{\eta}_{\bm{k}},\label{eom2}
\eeq
\end{subequations}
where we have defined $D^{\eta} = 1 + e\bm{\Omega}^{\eta}_{\bm{k}}\cdot\bm{B}$ as the modification of the density of states in the phase space due to the presence of the Berry curvature $\bm{\Omega}^{\eta}_{\bm{k}}$, $\bm{E}^{\eta} = e\bm{E} + \eta\bm{\mathcal{E}}^{\el}$ is the sum of the external electric field $\bm{E}$ and a pseudo electric field $\bm{\mathcal{E}}^{\el}=\dot{\bm{\mathcal{A}}}^{\el}$,  and the velocity $\bm{v}^{\eta} = \bm{v}^{0}_{\bm{k}} - e\partial_{\bm{k}}(\bm{m}^{\eta}_{\bm{k}}\cdot\bm{B})$, which is the sum of the group  velocity $\bm{v}^{0}_{\bm{k}}$ and a term coming from the presence of the orbital magnetic moment $\bm{m}^{\eta}_{\bm{k}}$. The presence of an orbital magnetic moment per valley might lead to a net macroscopic magnetization even in equilibrium if a different population in valleys is induced\cite{TCV05,XSN05,XYF06,XYN07}. Here we will describe the non-equilibrium situation.

The time evolution of the non equilibrium distribution function $f_{\eta}(t,\bm{k})$ will be described by the Boltzmann transport equation:

\beq
\tau\dot{f}^{\eta} + \tau\dot{\bm{k}}\cdot\partial_{\bm{k}}f^{\eta} = f_{0} - f^{\eta},\label{Boltzmann1}
\eeq

where $f_0$ is the equilibrium Fermi distribution function. For simplicity, we have assumed that all the external fields are homogeneous (they do not depend on space), but we will allow for time dependent electric fields: $\bm{E}^{\eta}(t)\sim \bm{E}^{\eta} e^{\pm \ii\omega t}$. Also, we will consider only elastic scattering with impurities within the relaxation time approximation\cite{BGK54}, neglecting intervalley scattering.

In all what follows we will assume that, at zero temperature, the Fermi level crosses the conduction band.
Since we will consider homogeneous fields, the distribution function is not a function of $\bm{x}$, and is thus enough to consider the current in the local limit:

\beq
\bm{J}^{\eta} = \frac{e}{4\pi^{2}}\int \dd^{2}\bm{k} D^{\eta}\dot{\bm{x}}^{\eta} f^{\eta}.\label{current}
\eeq

To first order in the external electric fields, the out of equilibrium distribution function can be written as 

\beq
f^{\eta} = f_{0} + \frac{\partial f_{0}}{\partial{\varepsilon}} (g^{\eta}_{+}(\omega,\bm{k})e^{\ii\omega t} + g^{\eta}_{-}(\omega,\bm{k})e^{-\ii\omega t}),\label{expansion}
\eeq

with $g^{\eta}_{\pm}$ of order $\mathcal{O}(\bm{E})$. Inserting the previous expression in (\ref{Boltzmann1}) and noticing that $\partial_{\bm{k}}f_0=\frac{\partial \varepsilon^{0}}{\partial\bm{k}}\frac{\partial f_0}{\partial\varepsilon}=\bm{v}^0\partial_{\varepsilon} f_0$, we have, for each component $g^{\eta}_{\pm}$, \emph{up to first order in the magnetic field}:

\beq
D^{\eta}\alpha_{\pm}g^{\eta}_{\pm} + \tau e\bm{B}\cdot(\bm{v}^{0}\times\partial_{\bm{k}})g^{\eta}_{\pm} = - \tau\bm{E}^{\eta}\cdot\bm{v}^{0},\label{Boltzmann2}
\eeq

with the factor $\alpha_{\pm}=1\pm \ii\omega\tau$. In two dimensions, we will consider the magnetic field perpendicular to the sample, $\bm{B}=B_3\hat{\bm{z}}$. For two dimensional massive Dirac systems with the dispersion relation $\varepsilon_{\bm{k}}=\pm\sqrt{v^2\bm{k}^2+m^2}$, the group velocity is $\bm{v}^0=\frac{v^2}{\varepsilon_{\bm{k}}}\bm{k}$,
so the second term of the left hand side of (\ref{Boltzmann2}) can be written, in polar coordinates as
\beq
\frac{\tau}{\alpha_{\pm}} e\bm{B}\cdot(\bm{v}^{0}\times\partial_{\bm{k}}) = - e B_{3}\frac{\tau}{\alpha_{\pm}} \frac{v^{2}}{\varepsilon_{\bm{k}}}\frac{\partial}{\partial\theta}\equiv -\gamma_{\pm} \frac{\partial}{\partial\theta},
\eeq
with the dimensionless parameter $\gamma_{\pm}$ defined as $\gamma_{\pm}=\frac{eB_3 v^2\tau}{\alpha_{\pm}\varepsilon_{\bm{k}}}$, and $\frac{1}{D^{\eta}}\simeq 1 - e B_{3}\Omega^{\eta}_{3} + \dots$

The Boltzmann equation then can be written as

\beq
( 1 - \gamma_{\pm}\partial_{\theta})g^{\eta}_{\pm} = - \frac{\tau}{\alpha_{\pm}}( 1 - e B_{3}\Omega^{\eta}_{3} )\bm{E}^{\eta}\cdot\bm{v}^{0}.\label{Boltzmann3}
\eeq

Let us apply the Jones-Zener method to solve (\ref{Boltzmann3}). This method consist in formally inverting the differential operator $(1-\gamma_{\pm}\partial_{\theta})$:
\beq
\frac{1}{(1-\gamma_{\pm}\partial_{\theta})}
\equiv 1 + \sum_{n=1}^{\infty}(\gamma_{\pm})^{n}(\partial_{\theta})^{n}
\simeq 1 + \gamma_{\pm}\partial_{\theta}+...,
\eeq
up to first order in the magnetic field, since $\gamma_{\pm}\sim\mathcal{O}(B_3)$.
For isotropic Dirac systems, the Berry curvature does not depend on $\theta$, so the expression for $g^{\eta}_{\pm}$ is then

\beq
g^{\eta}_{\pm}\simeq - \frac{\tau}{\alpha_{\pm}}( 1 - e B_{3}\Omega^{\eta}_{3} ) \bm{v}^{0}\cdot\bm{E}^{\eta} - e B_{3}\frac{\tau^{2}}{\alpha^{2}_{\pm}} \frac{v^{2}}{\varepsilon_{\bm{k}}}(\partial_{\theta}\bm{v}^{0})\cdot\bm{E}^{\eta}.\label{Solboltzmann}
\eeq

Let us stress that this expression is up to first order in the electric and magnetic fields, and, importantly, the modification of the group velocity due to the orbital magnetic moment $\bm{m}^{\eta}_{\bm{k}}$ starts to enter at second order in the magnetic field.

The next step is to use eq.(\ref{Solboltzmann}) together with (\ref{expansion}) and(\ref{eom2}) in (\ref{current}), again, keeping linear terms in the magnetic and electric fields.

The first contribution that depends only of the Fermi surface is the contribution corresponding to the standard conductivity (when $\bm{E}^{\eta} = e\bm{E}$) and it is independent of the magnetic field:

\beq
\bm{J}^{\eta,(1)} = - e\frac{1}{4\pi^{2}}\frac{\tau}{\alpha_{\pm}}\int \dd^{2}\bm{k}\frac{\partial f_{0}}{\partial\varepsilon} \bm{v}^{0} (\bm{v}^{0}\cdot\bm{E}^{\eta}).
\eeq

If we consider an elastic field $\bm{E}^{\eta} = \eta\dot{\bm{\mathcal{A}}}^{\el}$, noticing that the group velocity is the same for both valleys, we see that this part of the current acquires an opposite sign for each valley and the net contribution is zero.
The first contribution depending on the magnetic field is the classical contribution to the Hall current:

\beq
\bm{J}^{\eta}_{H} = \frac{e^{2}}{4\pi^{2}}\frac{\tau^{2}}{\alpha^{2}_{\pm}}\int \dd^{2}\bm{k}\frac{\partial f_{0}}{\partial \varepsilon}\bm{v}^{0}\bm{B}\cdot(\bm{v}^{0}\times\partial_{\bm{k}}\bm{v}^{0}\cdot\bm{E}^{\eta}).
\eeq

As before, if the electric field corresponds to the elastic pseudoelectric field $\bm{\mathcal{E}}^{\el}$, each contribution has opposite sign to the other and the net current is zero, even in presence of a magnetic field. 

The last contribution to the current at first order in the magnetic field is

\beq
\bm{J}^{\eta,(2)} = \frac{e^{2}}{4\pi^{2}}B_{3}\frac{\tau}{\alpha_{\pm}}\int\dd^{2}\bm{k}\frac{\partial f_{0}}{\partial\varepsilon} \left(\partial_{\bm{k}}m^{\eta}_{3} + \Omega^{\eta}_{3}\bm{v}^{0}\right)\bm{v}^{0}\cdot\bm{E}^{\eta}.\label{mainexpression}
\eeq

This piece of the current appears due to the presence of a Berry curvature, and an orbital magnetic moment. Because under time reversal invariant conditions, each valley contributes with both quantities having exactly opposite sign, in presence of a real electric field the total current coming from this pieces cancels out. However, if we consider the elastic vector field, the two contributions sum up and we have

\beq
\bm{J}^{(2)} = \frac{e^{2}}{2\pi^{2}}B_{3}\frac{\tau}{\alpha_{\pm}}\int\dd^{2}\bm{k}\frac{\partial f_{0}}{\partial\varepsilon} \left(\partial_{\bm{k}}m_{3} + \Omega_{3}\bm{v}^{0}\right)\bm{v}^{0}\cdot\dot{\bm{\mathcal{A}}}^{\el}.\label{Suma_en_Valles}
\eeq

Using the expressions for $m_{3}$ and $\Omega_{3}$ for massive Dirac fermions (see Appendix \ref{appendix:geometrical}), we get

\beq
\bm{J}^{(2)} = \frac{3e^{2}}{4\pi}B_{3}\frac{\tau}{\alpha_{\pm}}\Theta(\mu-m)\frac{m(\mu^{2} - m^{2})}{\mu^{4}}v^{2}\dot{\bm{\mathcal{A}}}^{\el}.\label{Magnetopiezo}
\eeq

We note that, in contrast to the piezoelectric contribution (see Eq.\ref{Piezo} in the Appendix), here the induced current is parallel to the vector field $\dot{\bm{\mathcal{A}}}^{\el}$.
\begin{figure}
\includegraphics[scale=0.4]{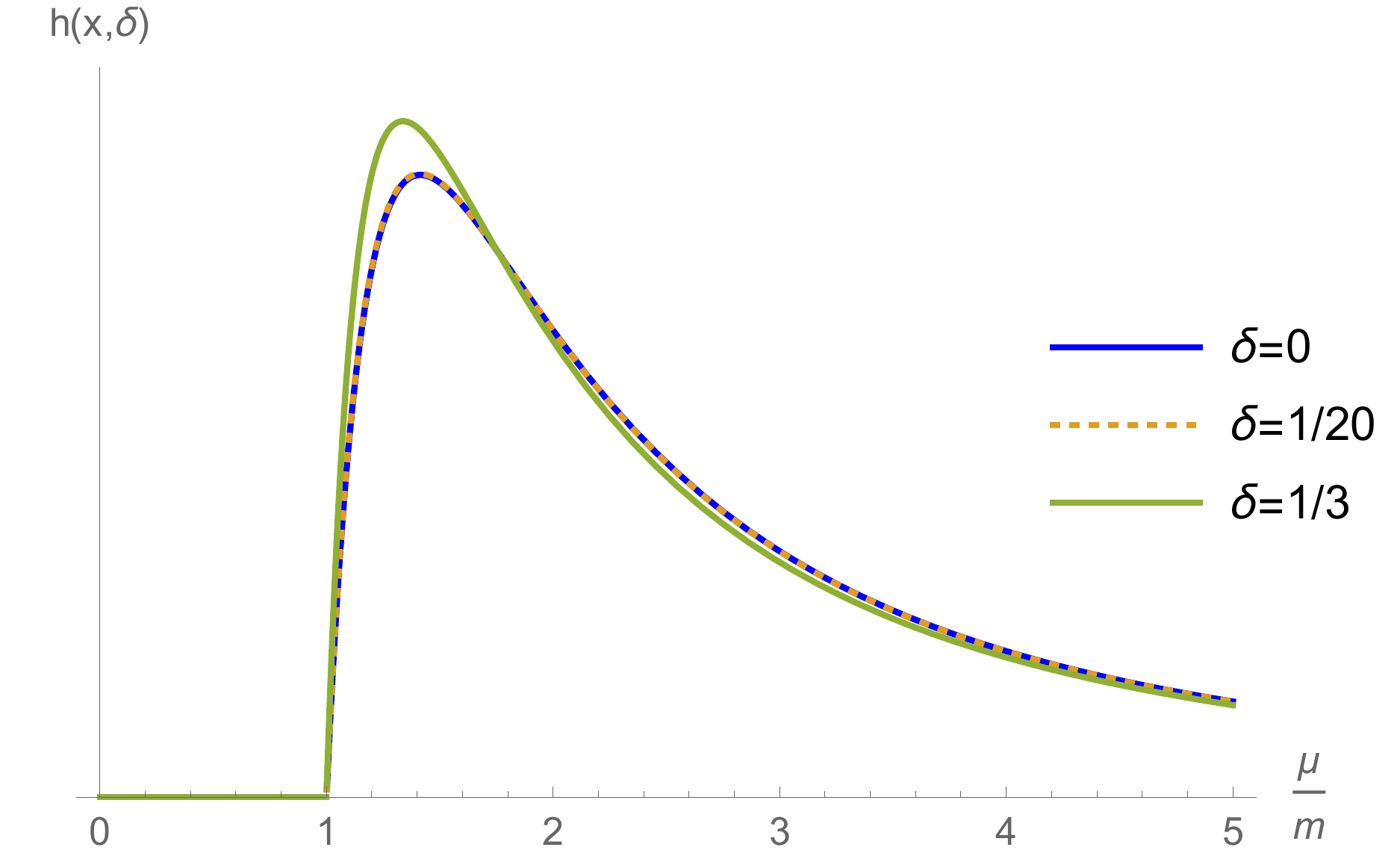}
\caption{(Color online) Behavior of the function  $h(x,\delta)$ ) of the magnetoelectric tensor with the chemical potential normalized to the mass parameter $m$ for massive Dirac, $\delta=0$ (blue solid line), MoS$_2$ (orange dashed line) with $\delta=1/20$, and for comparison, the case with $\delta=1/3$ (green solid line).}
\label{fig:magnetopiezo}
\end{figure}
This is the result we were looking for. In presence of time dependent elastic deformations, encoded in the pseudoelectric fied $\dot{\bm{\mathcal{A}}}^{\el}[u_{ij}]$, under the effect of an external magnetic field, a nonzero net electric current appears in the system. We note also that both the orbital magnetic moment through its derivative and the Berry curvature contribute to the current. For a single massive Dirac cone, breaking time reversal symmetry, this expression correspond to a linear magnetoconductivity\cite{XSN05}. Here the same expression defines the strain-induced magnetoelectric effect in time reversal invariant systems.

We have two regimes, depending on the effect of the disorder and the time dependence of the elastic deformations. As before, assuming harmonic dependence for the elastic deformations, we have $\dot{\bm{\mathcal{A}}}^{\el}\sim\pm \ii\omega  \bm{\mathcal{A}}^{\el}$
\beq
\bm{J}^{(2)} = \pm\frac{3e^{2}}{4\pi}B_{3}\frac{\ii\omega\tau}{\alpha_{\pm}}\Theta(\mu-m)\frac{m(\mu^{2}-m^{2})}{\mu^{4}}v^{2}\bm{\mathcal{A}}^{\el}.
\eeq
In the collisionless limit, $\omega\neq 0$ but $\omega\tau\gg 0$, so $\frac{\ii\omega\tau}{\alpha_{\pm}}\sim 1$ and the current $\bm{J}^{(2)}$ is proportional to $\bm{\mathcal{A}}^{\el}$:
\beq
\bm{J}^{(2)}\sim\frac{3e^{2}}{4\pi}B_{3}\Theta(\mu-m)\frac{m(\mu^{2}-m^{2})}{\mu^{4}}v^{2}\bm{\mathcal{A}}^{\el},
\eeq
while in the DC limit, corresponding to $\tau\neq 0$ but $\omega\to 0$, so $\frac{\ii\omega\tau}{\alpha_{\pm}}\sim \ii\omega\tau$ and 
\beq
\bm{J}^{(2)}\simeq\tau\frac{3e^{2}}{4\pi}B_{3}\Theta(\mu-m)\frac{m(\mu^{2}-m^{2})}{\mu^{4}}v^{2}\dot{\bm{\mathcal{A}}}^{\el}.
\eeq

In this limit we see that a DC electric polarization, rather than an electric current, is generated in the system under the effect of a time independent external strain, that is tuned by an external magnetic field, 
\beq
\bm{P}^{(2)}\simeq\tau\frac{3e^{2}}{4\pi}B_{3}\Theta(\mu-m)\frac{m(\mu^{2}-m^{2})}{\mu^{4}}v^{2}\bm{\mathcal{A}}^{\el}.\label{magnetoelectricpolarization}
\eeq

From the expression (\ref{magnetoelectricpolarization}), we can read off the non vanishing components of the strain-dependent magnetoelectric tensor $\alpha_{ij}$:

\begin{subequations}
\beq
\alpha^{D}_{13} = \tau\frac{3e^{2}}{4\pi}\frac{\beta}{a_{0}}\Theta(\mu-m)\frac{m(\mu^{2}-m^{2})v^{2}}{\mu^{4}}(u_{11}-u_{22}),
\eeq
\beq
\alpha^{D}_{23} = - \tau\frac{3e^{2}}{2\pi}\frac{\beta}{a_{0}}\Theta(\mu-m)\frac{m(\mu^{2}-m^{2})v^{2}}{\mu^{4}}u_{12}.
\eeq
\end{subequations}
\section{TMDCS}
So far, we have focused on the magnetoelectric tensor $\alpha_{ij}$ for massive Dirac systems (like BN) described by the model \ref{Dirac1}. We can obtain the tensor $\alpha_{ij}$ for TMDCs by using the expression (\ref{mainexpression}) and the Hamiltonian (\ref{DiracTMDCs}).

For the case of TMDCs, the previous expressions are a little more complicated. By defining the dimensionless parameters $x=\mu/m$ and $\delta=\lambda_{so}/m$ and summing over spin and valley degrees of freedom, the coefficients of the magnetoelectric tensor read
\begin{subequations}
\beqa
\alpha^{TMDC}_{13} = \frac{\tau v^{2}}{m} \frac{3e^{2}}{2\pi}\frac{\beta}{a_{0}}h(x,\delta)(u_{11}-u_{22}),
\eeqa
\beqa
\alpha^{TMDC}_{23} = - \frac{\tau v^{2}}{m} \frac{3e^{2}}{\pi}\frac{\beta}{a_{0}}h(x,\delta)u_{12}.
\eeqa
\end{subequations}
where we have defined the function $h(x,\delta)$ as
\beqa
& &h(x,\delta)=\Theta(x-1)(x-1)\cdot\nonumber\\
&\cdot&\left(\frac{(1-\delta)(1+x-2\delta)}{(x-\delta)^4}+\frac{(\delta+1)(1+x+2\delta)}{(x+\delta)^4}\right).
\eeqa

The behavior of $h(x,\delta)$ with the Fermi level $\mu/m$ is shown in Fig.(\ref{fig:magnetopiezo}) for massive Dirac fermions ($\delta=0$) and for TMDCs. The value $\delta=1/20$ for the ratio between the spin-orbit parameter and the symmetry breaking parameter $\delta=\lambda_{so}/m$  corresponds to MoS$_2$ and it has been extracted from \cite{RRC15}.
It is easy to see that, in the limit $\delta\to 0$, we recover the expressions for the massive Dirac model, $\alpha^{D}_{ij}$. Also, we observe in Fig.(\ref{fig:magnetopiezo}) that the magnetoelectric response is not too sensitive to the presence of the spin-orbit coupling in (\ref{DiracTMDCs}).

Then, as mentioned in Section \ref{sec:models}, these components of the magnetoelectric tensor will define a magnetization $M_3=\alpha_{3i}E_i=\alpha_{3i}\rho^{0}_{ij}J_{j}$, in terms of the resistivity tensor $\rho^{0}_{ij}$ and the applied current $\bm{J}$
\cite{PWD17}.
\section{Magnetic field-induced charge density}
We can explore other consequences of this strain-induced magnetoelectric effect, beyond the current experimental observations. For instance, if the strain depends on the space, instead of time, there is not induced current, but there is an induced inhomogeneous charge distribution 
$\rho_{b}(\bm{r})=-\bm{\nabla}\cdot\bm{P}^{(2)}(\bm{r})$ in the system that can be tuned by the external magnetic field whenever $\mu >m$:
\beqa\label{Magnetic_induced_charge_density}
\rho_{b}(\bm{r}) = - \tau\frac{3e^{2}}{4\pi} B_{3}\frac{m(\mu^{2}-m^{2})}{\mu^{4}}v^{2}\bm{\nabla}\cdot\bm{\mathcal{A}}^{\el}.
\eeqa
Using the expressions for the elastic field $\bm{\mathcal{A}}^{\el}$ in (\ref{Afield}), we have that the induced charge is
\beq
\rho_{b}(\bm{r})\sim \tau B_{3}(\partial_{1} u_{11} - \partial_{1} u_{22} - 2\partial_{2} u_{12}).
\eeq
Also, contrary to the piezoelectric contribution, the induced inhomogeneous density depends crucially on the presence of disorder in the sample.

\begin{figure}
\includegraphics[scale=0.6]{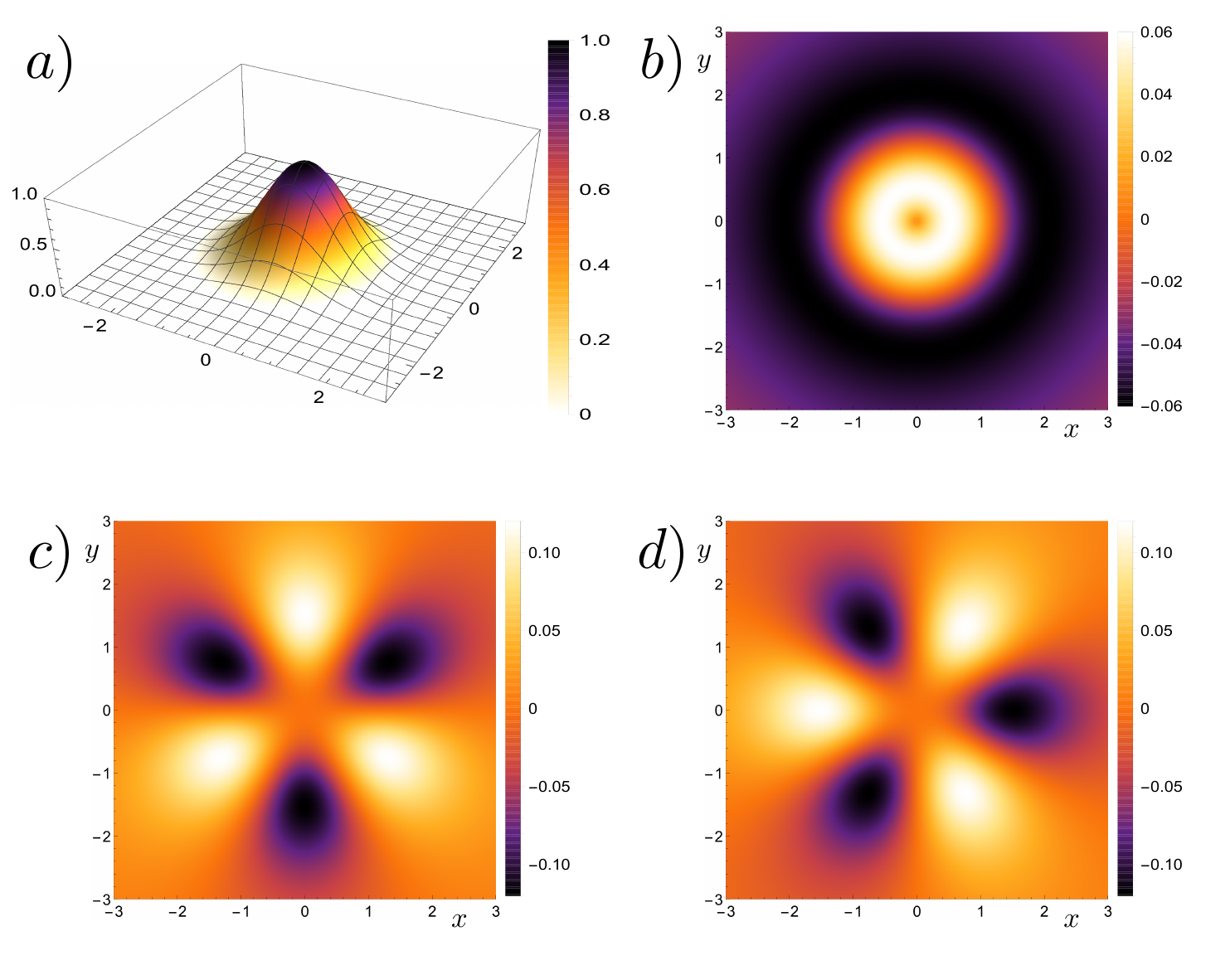}
\caption{(Color online) a) Three dimensional Gaussian deformation of a 2D material. b) In plane deformation of the 2D material induced by the Gaussian deformation. c) Piezoelectric induced charge density (divided by $\frac{e}{4\pi}$) and d) Magnetic field induced charge density (in units of $\tau\frac{3e^{2}}{4\pi}B_{3}\frac{3}{8m}v^{2}$), both cases for $\mu = 2m$.}
\label{fig:deformation_induced_charge_density}
\end{figure}

As an example, we study the effect of a Gaussian deformation present in the sample in Fig.~\ref{fig:deformation_induced_charge_density}a \cite{JCV07,JCV11}. For illustrative purposes we have chosen $\mu = 2m$. We obtain the in-plane induced deformation (Fig.~\ref{fig:deformation_induced_charge_density}b) by using the formalism developed in \cite{RGP17}. Then, a direct application of Eq.~\eqref{Definition_Gauge_Ael} in Eq.~\eqref{Magnetic_induced_charge_density}, leads to the result shown in Fig.~\ref{fig:deformation_induced_charge_density}c. We compare this result with the piezoelectric induced charge distribution, obtained from Eq.~\eqref{Piezoelectric_induced_charge_density} (Fig.~\ref{fig:deformation_induced_charge_density}d). We note that the induced charge density distributions are similar except for a multiplicative factor, and that they are rotated an angle $\theta = \frac{\pi}{6}$ from each other. This means that, in an in-homogeneously strained sample, the piezoelectric density distribution will be present, and it is independent of the magnetic field, whereas when an external magnetic field is switched on, the strain induced magnetoelectric charge density will appear.

For a Gaussian deformation of radius $R=0.5\mu m$ and with $\mu = 1.1m$, the piezoelectric induced charge density is $\rho_{p}\thickapprox 5\times 10^{10} e/cm^{2}$.
Using that mean free path of MoS$_{2}$ is $\ell = 4\times 10^{-8}m$\cite{BWW17}, under a magnetic field $B_{z} = 4T$, the magnetic field induced charge density is found to be $\rho_{b}\thickapprox 10^{9} e/cm^{2}$, roughly one order of magnitude smaller than the piezoelectric response, but still within the experimental range (see, for instance, \cite{EGM11}).

\section{Conclusions}

We have developed a theory on the strain-induced magnetoelectric coupling. Contrary to previous explanations, the magnetoelectric tensor $\alpha_{ij}$ is not related to the presence of piezoelectric currents, although the presence of a finite strain distortion is crucial for this magnetoelectric effect to appear. Also, its value is related to both the presence of the derivative of the orbital magnetic moment and a Berry curvature. We have estimated typical values for the charge density induced by inhomogeneous strains in the sample, of the order of $10^{9} e/cm^{2}$.
It is also worth to mention that this magnetoelectric effect is the two-dimensional counterpart of the magnetopiezoelectric effect recently described in three dimensional polar metals\cite{VGI17}.
\section{Acknowledgements}
A.C. acknowledges financial support through the MINECO/AEI/FEDER, UE Grant No. FIS2015-73454-JIN. and the Comunidad de Madrid MAD2D-CM Program (S2013/MIT-3007). 
P. R.-L. acknowledges partial support from TerMic (Grant No. FIS2014-52486-R, Spanish Government) and from Juan de la Cierva - Incorporacion program.
We acknowledge J. A. Garcia-Guillen for useful comments about TMDCs, and M. A. H. Vozmediano and P. Lopez-Sancho for a careful reading the manuscript.
\appendix
\section{Geometrical quantities}
\label{appendix:geometrical}
In this appendix we will present the expressions for the Berry curvature $\Omega_3$ and $m_{3}$ for the massive Dirac (Eq.(\ref{Dirac1})) and TMDCs (Eq.(\ref{DiracTMDCs})) models.
The general expression for the Berry curvature and the orbital magnetic moment are, for the band $n$:

\begin{subequations}
\beq
\Omega^{n}_{i} = i\varepsilon_{ijl}\braket{\partial_{j}n|\partial_{l}n},
\eeq
\beq
m^{n}_{i} = \frac{i}{2}\varepsilon_{ijl}\braket{\partial_{j}n|H(\bm{k})-\varepsilon^{\eta}_{\bm{k}}|\partial_{l}n}.
\eeq
\end{subequations}

In two dimensions for the conduction band ($n=+1$) of TMDCs, these expressions reduce to
\begin{subequations}
\beq
\Omega^{+}_{3}=-\eta\frac{v^2}{2\hat{\varepsilon}^3_{\bm{k}}}(m-s\eta \lambda_{so})\equiv\eta\Omega_3,
\eeq
\beq
m^+_{3}=\eta\frac{v^2}{2\hat{\varepsilon}^2_{\bm{k}}}(m-s\eta \lambda_{so})\equiv\eta m_3.
\eeq
\end{subequations}
\beq
m^+_{3}=-\hat{\varepsilon}_{\bm{k}}\Omega^{+}_{3}.
\eeq

In these expressions, $\hat{\varepsilon}_{\bm{k}} = \sqrt{v^{2}k^{2} + ( m - s\eta \lambda_{so})^{2}}$. The expressions for the massive Dirac case are recovered by setting $\lambda_{so}=0$.

\section{Piezoelectric response}
\label{appendix:piezo}
In this appendix we derive the piezoelectric response as the counterpart of the anomalous Hall contribution to the current for the effective vector field $\bm{\mathcal{A}}^{\el}$. We will refer all the time to quantities for the conduction band:

\beq
\bm{J}^{\eta}_{AH} = e\frac{1}{4\pi^{2}}\int \dd^{2}\bm{k}f_{0}\Omega^{\eta}_{3}(\hat{\bm{z}}\times\bm{E}^{\eta}),
\eeq

with $\epsilon_{ij}$ being the Levi-Civitta symbol in two dimensions. Contrary to other terms of the current, this piece is an \emph{equilibrium} contribution to the current since it is proportional to $f_0$. It also implies that all the occupied states contribute to this part of the current, in contrast to the other terms that only depend on the states close to the Fermi level (due to the dependence with $\partial_{\varepsilon}f_{0}$). Because of this fact, we need to be careful and consider the possibility that other bands can also contribute to this current, like the totally occupied valence band. For the simplistic case of massive Dirac fermions, the valence band, characterized by the Fermi distribution function $f^{v}_{0}$, has the same expression for the Berry curvature $\Omega^{\eta}_{3}$, but with opposite sign.

Using $\varepsilon=\pm\sqrt{v^2k^2+m^2}$, with the $\pm$ sign corresponding to conduction and valence bands, together with $\Omega^{\eta}_{3} = - \frac{\eta}{2}\frac{mv^{2}}{\varepsilon^{3}}$ we have ($\mu$ being the Fermi level):

\beq
\bm{J}^{\eta}_{AH} = \eta\frac{e}{4\pi}\frac{m}{\mu}\Theta(\mu-m)(\hat{\bm{z}}\times\bm{E}^{\eta}).
\eeq

This expression means that, due to the presence of a finite Berry curvature, a polarization current appears in the system, induced by an electric-like field $\bm{E}^{\eta}$. If we consider a real electric field $\bm{E}^{\eta} = e\bm{E}$, we need to break time reversal symmetry in order to lift the degeneracy of valleys imposed by $T$. However, if we consider an elastic electric field $\bm{E}^{\eta} = \eta\dot{\bm{\mathcal{A}}}^{\el}[u_{ij}]$ that couples with opposite sign to each valley, this sign cancels out with the sign coming from the Berry curvature at each valley, inducing an anomalous Hall like current in time reversal symmetric Dirac systems. The total Hall current is, summing over valleys:

\beq
\bm{J}_{AH} = \sum_{\eta=\pm}\bm{J}^{\eta}_{AH} = \frac{e}{2\pi}\frac{m}{\mu}\Theta(\mu-m)(\hat{\bm{z}}\times\dot{\bm{\mathcal{A}}}^{\el}).\label{Piezo}
\eeq

Using the standard expressions of $\bm{\mathcal{A}}^{\el}$ in terms of the strain tensor, we obtain the Berry phase-related piezoelectric response, after noticing that a current can be written as the time derivative of a polarization $\bm{P}$, $\bm{J}_{H}=\dot{\bm{P}}$:

\beq\label{Piezoelectric_induced_charge_density}
\bm{P} = \sum_{\eta=\pm}\bm{J}^{\eta}_{AH} = \frac{e}{2\pi}\frac{m}{\mu}\Theta(\mu-m)(\hat{\bm{z}}\times\bm{\mathcal{A}}^{\el}),
\eeq

plus contributions to the polarization that are not related to strain. It is important to stress that this piece of the current is independent of the magnetic field.


\end{document}